\documentclass[sigconf]{acmart}
\usepackage{subcaption}
\usepackage{booktabs} 
\usepackage{enumitem} 
\usepackage{makecell} 
\usepackage{algorithm,algorithmicx}
\usepackage[noend]{algpseudocode}
\usepackage{multirow}
\usepackage{multirow,bigdelim}
\usepackage{hhline}
\usepackage{pbox}
\usepackage{xspace}

\usepackage{graphicx}
\usepackage{color}
\usepackage{balance}
\usepackage{tikz}
\usetikzlibrary{calc}

\newcommand{\splitbrain}{\textsc{Splitter}\xspace}

\newcommand{\algrule}[1][.2pt]{\par\vskip.5\baselineskip\hrule height #1\par\vskip.5\baselineskip}
\algnewcommand{\LineComment}[1]{\State \(\triangleright\) #1}

\newcommand{\np}{\operatorname{np}}

\newcommand{\A}{{\mathcal A}}

\renewcommand{\comment}[1]{}

\setcopyright{rightsretained}

\copyrightyear{2019}
\acmYear{2019}
\setcopyright{iw3c2w3}
\acmConference[WWW '19]{Proceedings of the 2019 World Wide Web Conference}{May 13--17,
2019}{San Francisco, CA, USA}
\acmBooktitle{Proceedings of the 2019 World Wide Web Conference (WWW '19), May 13--17, 2019,
San Francisco, CA, USA}
\acmPrice{}
\acmDOI{10.1145/3308558.3313660}
\acmISBN{978-1-4503-6674-8/19/05}

\begin{document}
\title[Learning Node Representations that Capture Multiple Social Context]{Is a Single Embedding Enough? Learning Node Representations that Capture Multiple Social Contexts}

\author{Alessandro Epasto}
\orcid{0000-0003-0456-3217}
\affiliation{%
  \institution{Google AI}
  \city{New York}
  \state{NY}  
}
\email{aepasto@google.com}

\author{Bryan Perozzi}
\affiliation{%
  \institution{Google AI}
  \city{New York}
  \state{NY}
}
\email{bperozzi@acm.org}

\renewcommand{\shortauthors}{Epasto and Perozzi}

\begin{abstract}
Recent interest in graph embedding methods has focused on learning a single representation for each node in the graph.
But can nodes really be best described by a single vector representation?
In this work, we propose a method for learning multiple representations of the nodes in a graph (e.g., the users of a social network).  
Based on a principled decomposition of the ego-network, each representation encodes the role of the node in a different local community in which the nodes participate.
These representations allow for improved reconstruction of the nuanced relationships that occur in the graph -- a phenomenon that we illustrate through state-of-the-art results on link prediction tasks on a variety of graphs, reducing the error by up to $90\%$.
In addition, we show that these embeddings allow for effective visual analysis of the learned community structure.
\end{abstract}



\maketitle

\section{Introduction}
Learning embedded representations of graphs is a recent and very active area~\cite{deepwalk,line,walklets,node2vec,chen2017harp,asymmetric}. In a nutshell, an embedding algorithm learns a latent vector representation that maps each vertex $v$ in the graph $G$ to a {\it  single} $d$ dimensional vector. This area has found strong applications, as the embedding representation of nodes leads to improved results in data mining and machine learning tasks, such as node classification~\cite{deepwalk}, 
user profiling~\cite{Perozzi:2015:EAP:2740908.2742765}, ranking~\cite{hsu2017unsupervised}, and link prediction~\cite{node2vec,asymmetric}. 
In virtually all cases, the crucial assumption of the embedding methods developed so far is that a {\it single} embedding vector has to be learned for each node in the graph. Thus, the embedding method can be said to seek to identify the single role or position of each node in the geometry of the graph.

This observation allows us to draw a historical parallel between the very recent research area of graph embedding and the more established field of community detection or graph clustering~\cite{fortunato2010community,girvan2002community}. 
Detecting clusters\footnote{Note that in the paper, we use the terms ``cluster''
and ``community'' interchangeably.} in real world networks is a central topic in computer science, which has an extensive literature. At the beginning of its development, graph clustering has focused mostly on the study of non-overlapping clustering methods~\cite{suaris1988algorithm,girvan2002community}. In such methods, each node is assigned to a {\it single} cluster or community. While the problem of non-overlapping clustering is better understood and has found strong applications and theoretical results, recently, much attention has been devoted to developing {\it overlapping} clustering methods~\cite{EpastoKDD2017,crgp12,DBLP:conf/asunam/ReesG10}, where each node is allowed to participate in multiple communities. 
This interest in overlapping comunities is motivated by a number of recent observations of real world networks \cite{DBLP:journals/im/LeskovecLDM09,DBLP:conf/www/LeskovecLM10,crgp12,EpastoVLDB2016,ashk12} that show a lack of clear (non-overlapping) community structure.

These findings motivate the following research question: can embedding methods benefit from the awareness of the overlapping clustering structure of real graphs? In particular, can we develop methods where nodes are embedded in multiple vectors, representing their participation in different communities? 

In this paper, we provide positive results for these two questions. We develop \splitbrain, an unsupervised embedding method that allows nodes in a graph to have multiple embeddings to better encode their participation in multiple overlapping communities. Our method is based on recent developments in ego-net  analysis, in particular, in overlapping clustering algorithms based on ego-network partitioning~\cite{EpastoKDD2017}. 
More precisely, we exploit the observation in~\cite{crgp12,DBLP:conf/asunam/ReesG10,EpastoVLDB2016,EpastoKDD2017} that cluster structure is easier to identify at the local level.
Intuitively, this happens because each node interacts with a given neighbor in usually a single context (even if it is part of many different commmunities in total). 

\splitbrain extends this idea to the case of node embeddings. 
In particular, we exploit the {\it persona graph} concept defined by Epasto et al.~\cite{EpastoKDD2017}. This method, given a graph $G$, creates a new graph $G_P$ (called the persona graph of $G$), where each node $u$ in $G$ is  represented by a series of replicas.
These replica nodes, (called the \emph{persona}(s) of $u$) in $G_P$, represents an instantiation of the node $u$ in the local community to which it belongs. 
The method was originally introduced to obtain overlapping clusters. 
In this paper, we instead show that ego-net based techniques can lead to improvements in embedding methods as well. In particular, we demonstrate that a natural embedding method based on the persona graph outperforms many embedding baselines in the task of link prediction.

To summarize, the contributions of this paper are as follows:
\begin{enumerate}
 \item We introduce \splitbrain, a novel graph embedding method that embeds each node in the graph into multiple embedding vectors, which is based on the analysis of the overlapping structure of communities in real networks.
 \item Our method adds a novel graph regularization constraint to the optimization that enforces consistency of the multiple learned representations for each node.
 \item The method automatically determines the number of embeddings used for each node depending on a principled analysis of the local neighborhood of the node. This does not require the user to specify the number of embeddings as a parameter. 
 \item We show experimentally strong improvements over several embedding baselines for the important task of link prediction.
 \item We show how our method enables visual discovery of community membership for nodes that are part of multiple social groups.
\end{enumerate}

\section{Method}

In this section we describe  \splitbrain, our proposed method for learning multiple community-aware representations for a node.  
First, we start with a review of the preliminaries necessary to understand the work in Section \ref{sec:preliminaries}.  
Next, in Section  \ref{sec:multiple_representations} we discuss our extension for learning multiple node representations.
Then we introduce \splitbrain in Section \ref{sec:splitter}, and close with discussing some details of the optimization in Section \ref{sec:optimization}.

\subsection{Preliminaries}
\label{sec:preliminaries}

Our method builds upon recent work related to node decomposition \cite{EpastoKDD2017}, and learning node representations with neural networks \cite{deepwalk,walklets,perozzi2016local}.
Here we describe the basics of both methods.

\subsubsection{Notation}
We begin with some notation.
Let $G = (V, E)$ be an undirected graph\footnote{The method can be also defined for directed graphs in the obvious way; however, we describe it for undirected graphs for simplicity} consisting of a set $V$ of nodes and a set $E \subset V \times V$ of edges.
Let $G[U] = (U, E \cap U \times U)$ be the induced graph of a subset of $G$'s nodes, $U \subset V$.
Given a node $u \in V$, we denote its neighborhood as the set of nodes connected to it
$N_u = \{ v; (u,v) \in E\}$, and its \emph{ego-network} (or ego-net) as the graph induced on the neighborhood $G[N_u]$.  
We note that the ego-net of $u$ does not include the node $u$ itself in this definition.
Finally, let $\A$ be a \emph{non-overlapping clustering} algorithm that given $G$ as an input, returns a partition $\A(G) = (V_1, \hdots, V_t)$ of the vertices $V$ into $t$ disjoint sets (let $\np_\A(G) = t$ denote the number of partitions in output.).

\subsubsection{Persona Decomposition}
\label{sec.persona}

The topic of community detection and graph clustering has been of great interest to the community over the last several decades.
While much work has focused on finding large clusters, it has been noted that while the community detection problem is hard at scale (the  \emph{macroscopic} level), it is relatively trivial when viewed locally (the \emph{microscopic} level) \cite{coscia2014uncovering,EpastoVLDB2016}.  
Using this intuition, a recent proposal from \citet{EpastoKDD2017} uses the clusters found in the ego-network of a node (its neighbors and their induced subgraph) as the basis to define a new graph, the \emph{persona graph}, $G_P$.
The nodes in this new graph, which are called \emph{personas}, divide the interactions of each original node in $G$ into several semantic subgroups, which capture different components (or senses) of its network behavior.

More formally, let us assume that we are given a graph $G$ and a clustering algorithm $\A$.  The persona decomposition (as proposed in \cite{EpastoKDD2017}) employs the following algorithm \textsc{PersonaGraph$(G,\A)$} to transform $G$ to its persona graph $G_P$:
\begin{enumerate}[leftmargin=*]
	\item For each node $u \in V$, we use the clustering algorithm $\A$  to
	partition the ego-net of $u$. Let $\A(G[N_u]) = \{ N_u^1, N_u^2, \hdots,
	N_u^{t_u}\}$, where $t_u = \np_\A( G[N_u])$.
	\item Create a set $V'$ of personas. Each node $v_o$ in $V$ will
	correspond to $t_{v_o}$ personas (the number of splits of the ego-net of $v_o$) in $V_P$, denoted by $v_i$ for $i = 1, \hdots,
	t_{v_o}$.
	\item Add edges between personas. If $(u,v) \in E$, $v \in
	N_u^i$ and $u \in N_v^j$, then add an edge $(u_i, v_j)$ to $E_P$.
\end{enumerate}

After using this procedure, one obtains the persona graph $G_P$ which has some interesting properties.  
First, every node in $G_P$ is a node from the original graph, split into one or more personas.
However, there is no additional connectivity information -- the number of edges in $G_P$ is equal to the number of edges in the persona graph.
This means that the space required to store $G_P$ is (almost) the same as the original graph.
Second, each node in the original graph can be mapped to its corresponding persona(s).
However, the community structure of $G_P$ can be wildly different from the original graph.  
Standard clustering methods, when run on $G_P$ instead genereate overlapping clusterings of $G$.
This phenomena of exposing differing clustering information is visualized further in Section \ref{sec:visualization}.

\subsubsection{Graph Embedding}
\label{sec:embedding}
Before introducing our method for learning multiple embeddings for each node, we first review the standard setting of network representation learning, in which a single embedding is learned for each node. The purpose of network embedding is to learn a mapping $\Phi \colon v \in V \mapsto \mathbb{R}^{|V|\times d}$, which encodes the latent structural role of a node in the graph. This, in practice, can be achieved by representing the mapping $\Phi$ as a $|V| \times d$ matrix of free parameters that are learned by solving an optimization problem.
Perozzi et al.~\cite{deepwalk} first introduced a modeling of the vertex representation that encodes the node as a function of its co-occurrences with other nodes in short truncated random walks.

More precisely, the method consists of performing multiple random walks over the social graph from each node. The sequences of these walks are then used to extract the co-occurrences of nodes in short sub-windows. These co-occurrences capture the diffusion in the neighborhood around each vertex in the graph, and explore the local community structure around a node.
More concretely, the goal of the embedding method is to learn a representation that enables an estimate of the likelihood of a vertex $v_i$ co-occurring with other nodes in the sub-window of a short random walk:
\begin{equation}
\Pr\Big(v_i \mid \big(\Phi(v_1), \Phi(v_2), \cdots, \Phi(v_{i-1})\big)\Big)
\end{equation}
Notice that the exact computation of this conditional probability is computationally expensive for increasing lengths of the random walks, so DeepWalk uses two techniques to address this challenge. First, the order of the neighboring vertices is ignored. Second, the method reverses the learning task; instead of predicting a missing vertex using the context, it addresses the opposite problem of predicting its local structure using the vertex itself.

These modifications result in the following optimization problem for computing the vertex representations of each node in DeepWalk: 
\begin{equation}
\begin{aligned}
& \underset{\Phi}{\text{minimize}}
& -\log \Pr \big(\{v_{i-w}, \cdots, v_{i+w}\} \setminus v_i \mid \Phi(v_i) \big) \\
\end{aligned}	
\label{deepwalk:eq:objective}
\end{equation}
In the DeepWalk~\cite{deepwalk} model, the probability of a node $v_{i}$ co-occurring with $v_{j}$ is estimated by using a softmax to map the pairwise similarity to a probability space,
\begin{equation}
Pr({v_{i}|v_{j}})= \frac{exp(\Phi(v_{i}) \cdot \Phi^\prime(v_{j}))}{\sum_{j\in\mathcal{V}}{exp(\Phi(v_{i}) \cdot \Phi^\prime(v_{j})}}
\label{eq:skipgram}
\end{equation}

\noindent 
Where $\Phi^\prime(v_{i})$ and $\Phi^\prime(v_j)$ represent the "input" and "output" embeddings for node $v_{i}$ and $v_{j}$ respectively \cite{word2vec}.

\subsection{Learning Multiple Node Representations}
\label{sec:multiple_representations}
As discussed so far, network representation learning seeks to learn a function that maps each node to its own representation.
Here, we discuss our modifications that were made in light of the fact that we wish to learn one or more representation for each node.

Using the persona decomposition discussed in Section \ref{sec.persona}, we can convert the input graph $G$ into the persona graph $G_P$.
From here, it seems like a straightforward application of existing methods to learn one representation for each node $v \in |V_P|$, and as such, learn one or more representation for each original node $u \in |V|$.  Unfortunately, this is not the case.
The strength of the persona decomposition is also a weakness - it can create a graph that is quite different from that of the original input.  
In fact, the persona graph can be so different that it may consist of many disconnected components, even if the original graph was connected!
Thus, these disconnected components can cause difficulties for representation learning methods.
To address these challenges, we propose two improvements to the learning algorithm.

First, we propose adding a constraint, that the persona representations be able to predict their original node in addition to predicting the personas around it in $G_P$.
Specifically, given a persona $v_i$, we propose to require its representation include a dependency on the node $v_o$ in the original graph $G$:
\begin{equation}
\begin{aligned}
& \Pr \big( v_o \mid \Phi_{G_P}(v_i) \big).
\end{aligned}
\label{persona_math}
\end{equation}

\noindent To control the strength of our graph regularization, we introduce the parameter $\lambda$, which combines with Eq.\ \eqref{deepwalk:eq:objective} to yield the following optimization problem: 
\begin{equation}
\begin{aligned}
& \underset{\Phi_{G_P}}{\text{minimize}}
& -\log \Pr \big(\{v_{i-w}, \cdots, v_{i+w}\} \setminus v_i \mid \Phi_{G_P}(v_i) \big) \\
&& -\lambda \log \Pr \big( v_o \mid \Phi_{G_P}(v_i) \big). \\
\end{aligned}	
\label{walklets:eq:objective}
\end{equation}

\noindent Put another way, this change to the optimization enforces that there are invisible edges to each persona's parent, informing the learning process and regularizing the degree to which the persona representations can deviate.  
This effectively lets the model reason about the different connected components that may exist after the persona transformation.
We note that in practice, we achieved good results on all graphs we considered by simply setting $\lambda=0.1$.

Secondly, we propose to also make the representation $\Phi_{G_P}(v)$ of a node $v$'s personas depend on its original representation $\Phi_G(v)$ as a prior via initialization.  Initializing all personas to the same position in $\mathbb{R}^{d}$, combined with the regularization term from Eq.\ \eqref{walklets:eq:objective}, constrains the persona embeddings to behave like cohesive parts of a single entity.
We note that this does not mean that all of a node's personas end up in the same position at the end of the optimization!  Instead, we find that personas with similar roles stay close together, while personas with different roles separate.  This is discussed further in Section \ref{sec:co-authorship-visualization}, where we examine a visualization of \splitbrain\ embeddings.
Finally, there is an additional benefit of using the representation of the original graph $\Phi_G$ as an initialization -- it can help avoid potentially bad random initializations, which can lower task performance \cite{chen2017harp}.

\paragraph{Inference with Multiple Representations}
Notice that our method outputs multiple embeddings for each node. To do inference of node features or to predict edges between nodes, one can use standard ML methods to learn a function of the multiple embeddings of each node (or pair of nodes). However, basic aggregations can work too.  In our experiments with link prediction, we simply use the maximum dot product over all pairs of embeddings of $u$ and $v$ to predict the likelihood of the pair of nodes being connected.

\subsection{\splitbrain}
\label{sec:splitter}
Using the ideas discussed so far, we present the details of our approach.

\begin{algorithm}[t!]
	\caption{\textsc{\splitbrain}. Our method for learning multiple representations of nodes in a graph} 
	\label{alg:splitbrain} 
	\begin{algorithmic}[1]
		\Require{}
		\Statex{$G(V,E)$, a graph}
		\Statex{$w$, window size}
		\Statex{$d$, embedding size}
		\Statex{$\mathcal{\gamma}$, walks per vertex}
		\Statex{$t$, walk length}
		\Statex{$\alpha$, learning rate}		
		\Statex{$\lambda$, graph regularization coefficient}
		\Statex{$\A$, clustering algorithm for the ego-nets}
		\Statex{\textsc{EmbedFn}, a graph embedding method which uses the dot-product similarity (e.g.\ DeepWalk, node2vec)}		
		\Ensure{}
		\Statex{$\Phi_{G_P}$ a matrix with one or more representations for each node}
		\Statex{$P2N$, a mapping of the rows of $\Phi_{G_P}$ to $V$ (the original nodes)}
		\algrule
		\Function{\splitbrain}{$G$, $\textsc{EmbedFn}$}
		\State{$G_{P} \leftarrow \textsc{PersonaGraph}(G,\A)$}
		\Comment{Create the persona graph $G_P$ of $G$ using clustering algorithm $\A$ and method in~\cite{EpastoKDD2017}. }
		\State{$P2N \leftarrow \emptyset$}
		\State{$\Phi_{G} \leftarrow \textsc{EmbedFn}(G, w, d, \gamma, t)$}
		\Comment{Embed original graph}	
		\For{{\textbf{each} $v_o \in V$}}
			\For{\textbf{each} $v_{j} \in \textrm{personas of } v_o $}				
				\State{$\Phi_{G_P}(v_{j}) \leftarrow \Phi_{G}(v_o)$}
				\Comment{Initialize $j$-th persona of $v_o$}
				\State{$P2N(v_j) \leftarrow v_o$}
			\EndFor					
		\EndFor			
	
		\For{$i=0$ to $\mathcal{\gamma}$}
			\State	$\mathcal{O} = \text{Shuffle}(V_{G_P})$
	        \For{\textbf{each} $v_i \in \mathcal{O}$}
				\State $\mathcal{W}_{v_i} = RandomWalk(G_P, v_i, $t$) $		
			    \For{\textbf{each} $v_j \in \mathcal{W}_{v_i}$}
					\For{\textbf{each} $u_k \in \mathcal{W}_{v_i}[j-w: j+w]$}
							\State  $J_{G_P}(\Phi_{G_P})_ = - \log{\Pr(u_k \mid \Phi(v_j))}$
							\State  $J_{G}(\Phi_{G_P})_ = - \log{\Pr(P2N(v_j)} \mid \Phi(v_j))$
				        	\State $\Phi_{G_P} = \Phi_{G_P} - \alpha * \Big( \frac{\partial J_{G_P}}{\partial \Phi_{G_P}}  + \lambda \frac{\partial J_{G}}{\partial \Phi_{G_P}} \Big)$
					\EndFor
				\EndFor		
			\EndFor						
		\EndFor			
		\State{\Return $\Phi_{G_P}$}, $P2N$
		\EndFunction
	\end{algorithmic}
\end{algorithm}

\subsubsection{Parameters}
In addition to an undirected graph $G(V,E)$, the clustering algorithm $\A$ used to obtain the persona graph, as well as the dimensionality of the representations $d$, our algorithm uses a number of parameters that control the embedding learning process. 
The first group of parameters deal with sampling $G$, an essential part of any graph embedding process.
Without loss of generality, we describe the parameters to control the sampling process in the notation of Perozzi et. al \cite{deepwalk}.
Briefly, they are $w$, the window size to slide over the random walk, $t$, the length of the random walk to sample from each vertex, and $\mathcal{\gamma}$ the number of walks per vertex.
We emphasize that our approach is not limited to simple uniform random walk sampling - any of the more recently proposed graph sampling strategies \cite{perozzi2016local,walklets,node2vec} can be applied in the \splitbrain\ model.

The next group of parameters control the optimization.  The parameter $\alpha$ controls the learning rate for stochastic gradient descent, and $\lambda$ effects how strongly the original graph representation regularizes the persona learning.

Finally, an embedding function \textsc{EmbedFn} is used to learn a representation $\Phi_G$ of the nodes in the original graph.
In order to use the learning algorithm that we specify, this embedding method simply needs to produce representations where the dot-product between vectors encodes the similarity between nodes.
The most popular graph embedding methods meet this criteria, including DeepWalk~\cite{deepwalk}, LINE~\cite{line}, and node2vec~\cite{node2vec}.

\subsubsection{Algorithm}
Here, we describe our full algorithm, shown in Algorithm~\ref{alg:splitbrain} in detail.
Lines 2-4 initialize the data structures, create the persona graph, and learn the embedding of the underlying graph.  
 We note that not all of the nodes will necessarily be split; this will depend on each ego-net's structure, as described in Section~\ref{sec.persona}.
Lines 5-8 use the persona graph to initialize the persona representations $\Phi_{G_P}$.
The remainder of the algorithm (lines 9-17) details how the persona representations are learned.
Line 12 generates the random walks to sample the context of each vertex.
WLOG can be graph samples generated in any meaningful way - including uniform random walks~\cite{deepwalk}, skipped random walks~\cite{walklets}, or random walks with backtrack~\cite{node2vec}.
Line 15 calculates the loss due to nodes in the persona graph (how well the persona representation of $\Phi_{v_j}$ is able to predict the observed node $u_k$).
Line 16 computes the loss due to the graph regularization (how well the persona representation $\Phi_{v_j}$ is able to predict its corresponding original node).
Finally, line 18 returns the induced representations $\Phi_{G_P}$ and their mapping back to the original nodes $P2N$.

\paragraph{Complexity}
The complexity of the algorithm is dominated by two major steps: creating the persona graph and the Skip-gram model learning. Both parts have been analyzed in previous works~\cite{chen2017harp,EpastoKDD2017}, so we report briefly on the complexity of our algorithm here. Suppose the clustering algorithm $\A$ used on the ego-nets has a complexity of $T(m')$ for analyzing an ego-net of $m'$ edges. Further, suppose the original graph has $\mathcal{T}$ triangles. As such, the persona creation method has a total running time $O(\mathcal{T} + m + \sqrt{m} T(m))$. Moreover, as a worst case, $\mathcal{T} = O(m^{3/2})$ the complexity is $O(m^{3/2} + \sqrt{m} T(m))$. Suppose, for instance, that a linear time clustering algorithm is used; then, the total cost of this phase as a worst case is $O(m^{3/2})$. However, as observed before~\cite{EpastoKDD2017}, the algorithm scales much better in practice than this pessimistic bound because the number of triangles is usually much smaller than $m^{3/2}$. 
The embedding method for a graph of $n'$ total nodes in the persona graph has instead a complexity $O(n'\gamma t w d \log(n')))$, as shown in~\cite{chen2017harp}, where $d$ is the number of dimensions, $t$ is the walk length size, $w$ is the window size and $\mathcal{\gamma}$ is the number of random walks used. Notice, that $n' \in O(m)$, as each node $u$ can have at most $O(deg(u))$ persona nodes, so the worst case complexity is: $O(m^{3/2} + \sqrt{m} T(m) + m\gamma t w (d + d \log(m)))$.

\subsection{Optimization}
\label{sec:optimization}
As detailed in \cite{deepwalk}, using representations to predict the probability of nodes (Line 15-16, Algorithm 1) is computationally expensive.
To deal with this, we use the hierarchical softmax \cite{word2vec} to calculate these probabilities more efficiently.
For completeness's sake, we remark that an alternative optimization strategy exists (using noise contrastive estimation \cite{word2vec}).

Thus, our model parameter set includes both $\Phi_{G_P}$ and  $T$, the parameters used internally by the tree in the hierarchical softmax.
Further, we use the back-propagation algorithm to estimate the derivatives in lines 15-16, and a stochastic gradient descent (SGD) to optimize the model's parameters.
The initial learning rate $\alpha$ for SGD is set to 0.025 at the beginning of the training and then decreased linearly with the number of nodes we have seen so far.
Additionally, the parameter $\lambda$ regularizes the persona embeddings by how much they deviate from the node's original representation.

\section{Task: Link Prediction}

In this section, we study how the \splitbrain\ model proposed so far can be used for the task of link prediction -- judging the strength of a potential relationship between the two nodes.  We focus on this fundamental application task for the following reasons.

First, we are interested in developing new models that can capture the variation of social behaviors that are expressed in real-world social networks (such as membership in multiple communities).
Several recent works suggest that link prediction (or \emph{network reconstruction}) \cite{asymmetric,Zhang:2018:APP:3219819.3219969} is the best way to analyze an unsupervised network embedding's performance, as it is a primary task (unlike node classification - a secondary task that involves a labeling process that may be uncorrelated with the graph itself).
Second, there are several important industrial applications of link prediction on real networks (e.g.\ friend suggestion on a social network, product recommendation on an e-commerce site, et cera).

Finally, this task highlights a particular strength of our method.  
\splitbrain's ability to model the differing components of a node's social profile (its \emph{personas}) make it especially suitable for the task of link prediction.
This addresses a fundamental weakness of most node embedding methods, which effectively treat a node's representation as an average of its multiple senses -- a representation that may not make sense in continuous space.
Unlike previous work utilizing community information in embeddings \cite{cavallari2017learning,wang2017community,zheng2016node}, we aim to expose the nuanced relationships in a network by sub-dividing the nodes (not the macro-scale community relationships found by joining nodes together).

\subsection{Experimental Design}

\subsubsection{Datasets}
We test our \splitbrain method as well as other baselines on a dataset of five directed and undirected graphs. Our datasets are all publicly available: PPI is introduced~\cite{ppi, node2vec}, while the other datasets are from Stanford SNAP library~\cite{snap}. For each dataset, in accordance with the standard methodology used in the literature~\cite{node2vec,asymmetric}, we use the largest weakly connected component of the original graph. We now provide statistics for our dataset.

\noindent \textbf{Directed graphs}:
\begin{enumerate}[topsep=0cm,labelsep=0cm,align=left,noitemsep,nolistsep]
	\item {\bf soc-epinions}: A social network $|V| = 75,877$ and $|E| = 508,836$. Edges represent who trusts whom in the opinion-trust dataset of Epinions.
	\item {\bf wiki-vote}: A voting network  $|V| = 7,066$ and $|E| = 103,663$. Nodes are Wikipedia editors. Each directed edges represents a vote for allowing another user becoming an administrator.
\end{enumerate}
\noindent \textbf{Undirected graphs}:
\begin{enumerate}[topsep=0cm,labelsep=0cm,align=left,noitemsep,nolistsep]
	\item {\bf ca-HepTh}: Arxiv's co-author network of High Energy Physics Theory $|V| = 9,877$ and $|E| = 25,998$. Each edge represents co-authorship between two author nodes.
	\item {\bf ca-AstroPh}:   Arxiv's co-author network of Astrophysics, $|V| = 17,903$ and $|E| = 197,031$. Each edge represents co-authorship between two author nodes.
	\item {\bf PPI}: Protein-protein interaction graph, $|V| = 3,852$ and $|E| = 20,881$. This represents a natural dataset, where each node is a protein and there is an edge betwen two proteins if they interact (more details in~\cite{node2vec}).  
\end{enumerate}

\begin{table*}[t!]
	\centering
	\setlength\tabcolsep{3.0pt}
	
\begin{tabular}{ r@{\hskip 0.4cm} r || c | c | c | c | c | c || c | c | c | c | c | c | c}
& \textbf{Dataset} & \multicolumn{6}{c||}{\textbf{Non-Embedding Adjacency Methods}} & \multicolumn{7}{c}{\textbf{Embedding Methods}} \\ 
\hhline{~--------------}
&     &   \multicolumn{3}{c|}{Original graph} &\multicolumn{3}{c||}{Persona graph} & &  \multicolumn{5}{c|}{Embedding Baselines}    & \multicolumn{1}{c}{\textbf{Ours} } \\
& & \multirow{2}{*}{J.C.}  & \multirow{2}{*}{\makecell{C.N.}} & \multirow{2}{*}{\makecell{A.A.}} &  \multirow{2}{*}{J.C.}  & \multirow{2}{*}{\makecell{C.N.}} & \multirow{2}{*}{\makecell{A.A.}}   &   d & \makecell{Eigen \\ Maps} & node2vec & DNGR  &  Asymmetric  & M-NMF  & \splitbrain  \\
\hhline{~==============}
\parbox[t]{2mm}{\multirow{10}{*}{\rotatebox[origin=c]{90}{directed}}\ldelim\{{10}{40pt}}
& \multirow{5}{*}{soc-epinions} & \multirow{5}{*}{0.649} & \multirow{5}{*}{0.649} & \multirow{5}{*}{0.647} & \multirow{5}{*}{0.797} & \multirow{5}{*}{0.797} & \multirow{5}{*}{0.797} & 8  & $\dagger$  & 0.725  & $\dagger$  & 0.695 & $\dagger$  &{\bf 0.972}\\ 
& & & & & & & & 16  & $\dagger$  & 0.726  & $\dagger$  & 0.699 &$\dagger$  & {\bf 0.974}\\ 
& & & & & & & & 32  & $\dagger$  & 0.714  & $\dagger$  & 0.700 &$\dagger$  &{\bf 0.973} \\ 
& & & & & & & & 64  & $\dagger$  & 0.699  & $\dagger$  & 0.698 &$\dagger$  &{\bf 0.970} \\ 
& & & & & & & & 128  & $\dagger$  & 0.691  & $\dagger$  & 0.718 &$\dagger$  &{\bf 0.967} \\ 

\cline{2-15}
& \multirow{5}{*}{wiki-vote} & \multirow{5}{*}{0.579} & \multirow{5}{*}{0.580} & \multirow{5}{*}{0.562} & \multirow{5}{*}{0.860} & \multirow{5}{*}{0.865} & \multirow{5}{*}{0.866} &  8  & 0.613  & 0.643  & 0.630  & 0.608 & 0.886&{\bf 0.950} \\ 
& & & & & & & & 16  & 0.607  & 0.642  & 0.622  & 0.643  & 0.912&{\bf 0.952}\\ 
& & & & & & & & 32  & 0.600  & 0.641  & 0.619  & 0.683 & 0.926& { \bf 0.953}\\ 
& & & & & & & & 64  & 0.613  & 0.642  & 0.598  & 0.702  &0.932 & {\bf 0.952}\\ 
& & & & & & & & 128  & 0.622  & 0.643  & 0.554  & 0.730  & 0.934& {\bf 0.939}\\ 

\hhline{~==============}
\parbox[t]{2mm}{\multirow{15}{*}{\rotatebox[origin=c]{90}{undirected}}\ldelim\{{15}{40pt}}
& \multirow{5}{*}{ca-HepTh} & \multirow{5}{*}{0.765} & \multirow{5}{*}{0.765} & \multirow{5}{*}{0.765} & \multirow{5}{*}{0.553} & \multirow{5}{*}{0.553} & \multirow{5}{*}{0.553} &  8  & 0.786  & 0.731  & 0.706  & 0.605 & 0.852& {\bf 0.877} \\ 
& & & & & & & & 16  & 0.790  & 0.787  & 0.780  & 0.885  & 0.884&{\bf 0.897}\\ 
& & & & & & & & 32  & 0.795  & 0.858  & 0.829  & 0.884  & 0.903&{\bf 0.909}\\ 
& & & & & & & & 64  & 0.802  & 0.886  & 0.868  & 0.870  & 0.912&{\bf 0.917}\\ 
& & & & & & & & 128  & 0.812  & 0.901  & 0.897  & 0.820 & 0.908&{\bf 0.920} \\ 

\cline{2-15}
& \multirow{5}{*}{ca-AstroPh} & \multirow{5}{*}{0.942} & \multirow{5}{*}{0.942} & \multirow{5}{*}{0.944} &\multirow{5}{*}{0.874} & \multirow{5}{*}{0.874} & \multirow{5}{*}{0.874} &  8  & 0.825  & 0.811  & 0.852  & 0.592 &0.903 &{\bf 0.959} \\ 
& & & & & & & & 16  & 0.825  & 0.833  & 0.877  & 0.657 &0.935 &{\bf 0.972} \\ 
& & & & & & & & 32  & 0.825  & 0.899  & 0.917  & 0.942 &0.954 &{\bf 0.978} \\ 
& & & & & & & & 64  & 0.824  & 0.934  & 0.939  & 0.936 &0.966&{\bf 0.982} \\ 
& & & & & & & & 128  & 0.829  & 0.955  & 0.968  & 0.939 &0.974&{\bf 0.985 }\\

\cline{2-15}
& \multirow{5}{*}{PPI} & \multirow{5}{*}{0.766} & \multirow{5}{*}{0.776} & \multirow{5}{*}{0.779} &\multirow{5}{*}{0.698} & \multirow{5}{*}{0.701} & \multirow{5}{*}{0.702} & 8  & 0.710  & 0.733  & 0.583  & 0.550 &0.739 &{\bf 0.865} \\ 
& & & & & & & & 16  & 0.711  & 0.707  & 0.687  & 0.786 &0.776 &{\bf 0.869}\\ 
& & & & & & & & 32  & 0.709  & 0.691  & 0.741  & 0.794 & 0.793&{\bf 0.869}\\ 
& & & & & & & & 64  & 0.707  & 0.671  & 0.767  & 0.813 & 0.817&{\bf 0.866} \\ 
& & & & & & & & 128  & 0.737  & 0.698  & 0.769  & 0.799 & 0.840&{\bf 0.863} \\ 
\end{tabular}
	\vspace*{3mm}
	\caption[Table caption text]{
		We report the ROC-AUC for a link prediction task performed using an ablation test. The columns J.C., C.N. and A.A. stand for the baselines jaccard coefficient, common neighbors and adamic-adar, respectively.
		
		The rows represents the datasets (directed and undirected) while the columns represent the methods compared. We compare our \splitbrain method with three non-embeddings methods (applied to both the original and persona graph) as well four embeddings baselines. For the embedding methods we report results using dimensionality  $\{8, 16, 32, 64, 128\}$. We report in \textbf{bold} the highest AUC-ROC for each dimension and dataset. Notice that for \splitbrain the dimension refers to the size of the embedding of each persona node (i.e. the total embedding size of each node is larger). The next table reports results at parity of space.
		Results with $\dagger$ indicate lack of completion. We used a machine with 32 GB ram.
	}
	\label{table:auc}
\end{table*}

\subsection{Task}
\label{sec.task.linkprediction}
The Link Prediction task follows the methodology introduced in \cite{node2vec,asymmetric}, which we briefly detail here.
First, the input graph is split into two edge sets, $E_\text{train}$ and $E_\text{test}$, of equal size.
The test edges are removed uniformly at random, with the restriction that they do not disconnect the graph.
$E_\text{test}$ is then used as positive examples for a classification task.
A corresponding equal sized set of non-existent (random) edges are generated to use as negative examples for testing.
The baseline methods are providing the training edges as input, which they use to learn a similarity model (embedded or otherwise).
The performance of each method is then measured by ranking the removed edges.  Specifically, for each method, we report the ROC-AUC.

\subsection{Methods}
Here we describe the numerous baseline methods we tested against, including both traditional non-embedding baselines (such as common neighbors) and several embedding baselines.  We also detail the application of \splitbrain's multiple representations of this task. \\

\noindent \textbf{Non-embedding Baselines}:
Here, we report standard methods for link prediction that are solely based on the analysis of the adjacency matrix of the graph and in particular, on the immediate neighborhood of the nodes in the graph.
These methods take into input $E_\text{train}$ during inference. Thus, we denote $N(u)$ as the neighbors of $u$ observed in $E_\text{train}$. For directed graphs, $N(u)$ only refers to the outgoing edges. 
In the non-embedding baseline considered, we score an edge $(u, v)$ as a $g(u,v)$, which is a function of $N(u)$ and $N(v)$ only. We consider the following baselines:
\begin{enumerate}[topsep=0cm,labelsep=0cm,align=left,noitemsep,nolistsep]
	\item \textbf{Jaccard Coefficient (J.C.) :} \[g(u, v)=\frac{|N(u) \cap N(v)|}{|N(u) \cup N(v)|}\]
	\item \textbf{Common Neighbors (C.N.):}  \[g(u, v) = |N(u) \cap N(v)|\]
	\item \textbf{Adamic Adar (A.A.):}	 \[g(u, v) = \sum_{x \in N(u) \cap N(v)} \frac{1}{\log(|N(x)|)} \]
\end{enumerate}
We apply these methods to both the original graph and the persona graph. For the persona graph, we follow a technique similar to our \splitbrain method to extract a single score from the pairwise similarity (say Jaccard Coefficient) of the multiple persona nodes of a pair of nodes $u$,$v$. We also define the Jaccard Coefficient of $u,v$ in the persona graph as the maximum  Jaccard Coefficient of a persona node of $u$ and a persona node of $v$ in the persona graph. Similarly, we define the baselines Common Neighbors and Adamic Adar in the persona graph. We report results for using the maximum as aggregation function consistently with our \splitbrain method, but we experimented as well with many other functions, such as the minimum and the mean and we observed the maximum to perform best.\\

\begin{table}[t!]
	\centering
	\setlength\tabcolsep{3.0pt}
	\begin{tabular}{ r | c  }
\textbf{Dataset} & \textbf{Avg. Personas per Node $\bar{p}$}  \\ 
\hhline{--}
soc-epinions & 3.03 \\
wiki-vote &  4.00 \\ 
ca-HepTh &  2.39\\
ca-AstroPh &  2.53\\
ppi &  4.97\\19159  
\end{tabular}
    \vspace*{3mm}
	\caption[Table caption text]{Average number of persona nodes per node in original graph.}
	\label{table:persona-per-node}
\end{table}

\noindent \textbf{Embedding Baselines}:
We also consider the following embedding baselines. These methods take as input $E_\text{train}$ to learn embedding $\Phi_G(u)$ for every graph node $u$. During inference, then only the learned embedding are used but \textit{not} the original graph. We compare against these state-of-the-art embedding methods:
\begin{enumerate}[topsep=0cm,labelsep=0cm,align=left,noitemsep,nolistsep]
	\item Laplacian \textbf{EigenMaps} \cite{eigenmaps} determines the lowest eigenvectors of the graph Laplacian matrix.
	\item \textbf{node2vec} \cite{node2vec} learns the embedding by performing random walks on the training edges $E_\text{train}$ and learning a minimizing skipgram objective (Equation \ref{eq:skipgram}).
	\item \textbf{DNGR} \cite{dngr} performs a non-linear (i.e. deep) node embeddings passing a ``smoothed'' adjancency matrix through a deep auto-encoder. The ``smoothing'' (called Random Surfing) an alternative to random walks, which effectively has a different context weight from node2vec.
	\item \textbf{Asymmetric} \cite{asymmetric} is a recent method that learns embeddings by explicitly modeling the edges that appear in the graph.  We compare against the most similar model proposed in the work, the shallow asymmetric model.
	\item \textbf{M-NMF} \cite{wang2017community} uses a modularity based community detection model to jointly optimize the embedding and community assignment of each node.  Unlike \splitbrain, this method assigns each node to one community, and is based on joining nodes together (not splitting them apart).	
\end{enumerate}
We run each of these methods with their suggested default parameters.\footnote{In order to advance the field, and ensure the reproducibility of our method, we are releasing an implementation of \splitbrain at \texttt{\url{https://github.com/google-research/google-research/tree/master/graph_embedding/persona}}.}  \\

During inference with the baselines, we use the embedding of a pair of node $u$ and $v$ to rank the likelihood of the link $u,v$ formed by employing a scoring function that takes in the input the embeddings of the two nodes. To do so, for consistency with previous work, we used the same methodology of \cite{asymmetric}, which we summarize here. Let $Y_u$ and $Y_v$ be, respectively, the embeddings of $u$ and $v$. The edge scoring function is defined as follows: for \textbf{EigenMaps}, it is $-||Y_u - Y_v||$; for \textbf{node2vec}, we use the off-shelve binary classification LogisticRegression algorithm of sklearn to lean a model over the Hadamard product of the embeddings of the two nodes; for \textbf{DNGR},  we use the bottleneck layer values as the embeddings and the dot product as similarity; for \textbf{Asymmetric}, we use the dot product; and for \textbf{M-NMF}, similarly to node2vec, we train a model on the Hadamard product of the embeddings.

\noindent \textbf{Our Method (\splitbrain)}:  In order to use \splitbrain\ for link prediction, we need a method to calculate a single similarity score between two nodes ($u$,$v$) in the original graph $G$, each of which may have multiple persona representations.
Specifically, as the \splitbrain\ embedding model uses the dot-product to encode the similarity between the two node's representations, we need a method to extract a single score from the pairwise similarity of the (potentially) multiple persona nodes.
Similar to applying non-embedding baselines to the persona graph, we experimented with a number of aggregation functions (including \textit{min}, \textit{max}, \textit{mean}, etc).
The highest performing aggregation function was the maximum, so we define the similarity between the two nodes in $G$ to be the maximum dot-product between any of their constituent personas in $G_P$.

For learning embeddings with \splitbrain, we set the random walk length $t$ = 40, number of walks per node $\gamma$ = 10, and the window size $w$ = 5, the initial learning rate $\alpha$ = 0.025, and the graph regularization coefficient $\lambda$ = 0.1.
For \textsc{EmbedFn}, we used node2vec with random walk parameters ($p=q=1$) which is equivalent to DeepWalk.

\subsection{Experimental Results}

\begin{table*}[t!]
	\centering
	\setlength\tabcolsep{3.0pt}
	\begin{tabular}{ r || c | c | c | c | c | c | c}
\textbf{Dataset} & $d_{max} = 16\bar{p}$ & \textbf{Best EigenMaps} & \textbf{Best Node2Vec}  & \textbf{Best DNGR}  & \textbf{Best Asymmetric} & \textbf{Best M-NMF} & \textbf{\splitbrain $d=16$}  \\
\hhline{========}
soc-epinions & 48.5 & $\dagger$  & 0.726 & $\dagger$ & 0.700 & $\dagger$  & {\bf 0.974} \\
\hline
wiki-vote &  64.0  & 0.613  &  0.643 & 0.630 &0.702 &0.932 & {\bf 0.952}  \\ 
\hline
ca-HepTh &  38.2 & 0.802 & 0.886 & 0.868 & 0.885 & {\bf 0.912} &  0.897 \\
\hline
ca-AstroPh &  40.5 & 0.824 &  0.934&0.939 & 0.942 &0.966 & {\bf 0.972} \\
\hline
ppi &  79.5 & 0.737 &0.733 &0.769 &0.813 &0.840 & {\bf 0.869} \\ 
\end{tabular}
	\caption[Table caption text]{AUC-ROC of \splitbrain with $d=16$ compared with best baseline allowed larger total embedding space (at approximate space parity).}
	\label{table:auc-same-space}
\end{table*}

In the following table, we report the experimental comparison of \splitbrain with several embedding and non-embedding baselines. For each experiment, we report the AUC-ROC in a link prediction task performed using an ablation test described in Section \ref{sec.task.linkprediction}. For consistency of comparison, we use the experimental settings (datasets and training/testing splits) of~\cite{asymmetric} for the baselines. Hence, the baselines’ numbers are the same of~\cite{asymmetric} and are reported for completeness. 

We first report in Table~\ref{table:auc} the results of \splitbrain with several dimensionality settings. All results involving \splitbrain or the adjacency matrix on persona graph baseline uses the connected component method for ego-net clustering. We chose the connected component algorithm for easy replication of the results because it performs very well, as well as due to its previous use in ego-net clustering~\cite{DBLP:conf/asunam/ReesG10,EpastoKDD2017}. In particular~\cite{EpastoKDD2017}, showed theoretical results in a graph model for the connected component method at ego-net level.

We first take a look at the adjacency matrix baselines. Here, we consider both the vanilla version of the well-known baselines in the original graph, as well as the application of such baselines on the persona pre-processed graph (with a methodology similar to \splitbrain, as described above). We observe that simply applying the persona preprocessing to the standard baselines does not consistently improve the results over using them in the original graph. In particular, we only observe improvements in two of the five graphs we analyzed, while sometimes, we see even strong losses in applying this simple pre-processing, especially for our sparsest graphs, such as ca-HepTh. This confirms that the gains observed in our \splitbrain do not come merely from the pre-possessing.

Now, we consider the embedding methods. In this table, to gain an understanding of \splitbrain embeddings, we compare different sizes of \splitbrain embeddings (per persona node) with same size embeddings of other methods (per node). Before delving into the results, a note is important; since each node can have multiple persona embeddings, the total embedding size of \splitbrain (for the same dimension) can be larger than that of another standard embedding method. For this reason, we will later compare the results at same total embedding size. First, we observe in Table~\ref{table:auc} that at the same level of dimensionality, \splitbrain always outperforms all other baselines. The improvement is particularly significant in the largest graph {\it epinions} where our method using size $8$ embeddings improves AUC-ROC by a factor of $40\%$ (reduction in error of $90\%)$ even when compared with the best baseline with 128 dimensions. Similarly, our method achieves close to optimal performances in two of the other largest graphs, wiki-vote and ca-AstroPh.

As we have mentioned before, our method embeds each node into multiple embeddings (one for each persona node), so for a given dimension $d$ of the embedding, the average embedding size per node is given by $d \bar{p}$, where $\bar{p}$ is the average number of personas per nodes (i.e. the average number of ego-net clusters per node) in the graph. We report the average number of persona nodes for all the graphs in our datasets in Table~\ref{table:persona-per-node}. As we notice, the average number of persona nodes is between $2$ and $5$ in our datasets (using the connected component ego-network splitting algorithm). We report in Table~\ref{table:auc-same-space} a different look at the previous results. This time, we compare \splitbrain's AUC-ROC to other embeddings, allowing for the same (or higher) total embedding space. In our example in Table~\ref{table:auc-same-space},  we fix $d=16$ for the \splitbrain method and then we compute the effective average embedding size for each dataset ($\bar{p} 16$). Thereafter, we compare our results with the best result for each baseline that uses approximately $16 \bar{p}$ dimensions (for fairness of comparison, we actually round $16\bar{p}$ to the next power of $2$ and always allow more space for the other methods vs our method). 
Thus, it is possible to observe that the AUC-ROC of \splitbrain 
in is higher than that of every other baseline using about $16\bar{p}$ dimensions for all datasets, except once (in ca-HepTh, M-NMF is better, and we observe this is the sparsest graph). This confirms that the method improves over the baselines even after accounting for the increased space due to the presence of multiple embeddings for nodes. We observe the same results for other $d$ besides $d=16$.

\section{Task: Visualization}
\label{sec:visualization}
\begin{figure*}[h!]
\centering
  \begin{subfigure}[t]{0.4\textwidth}
  \centering
    \includegraphics[width=0.5\textwidth]{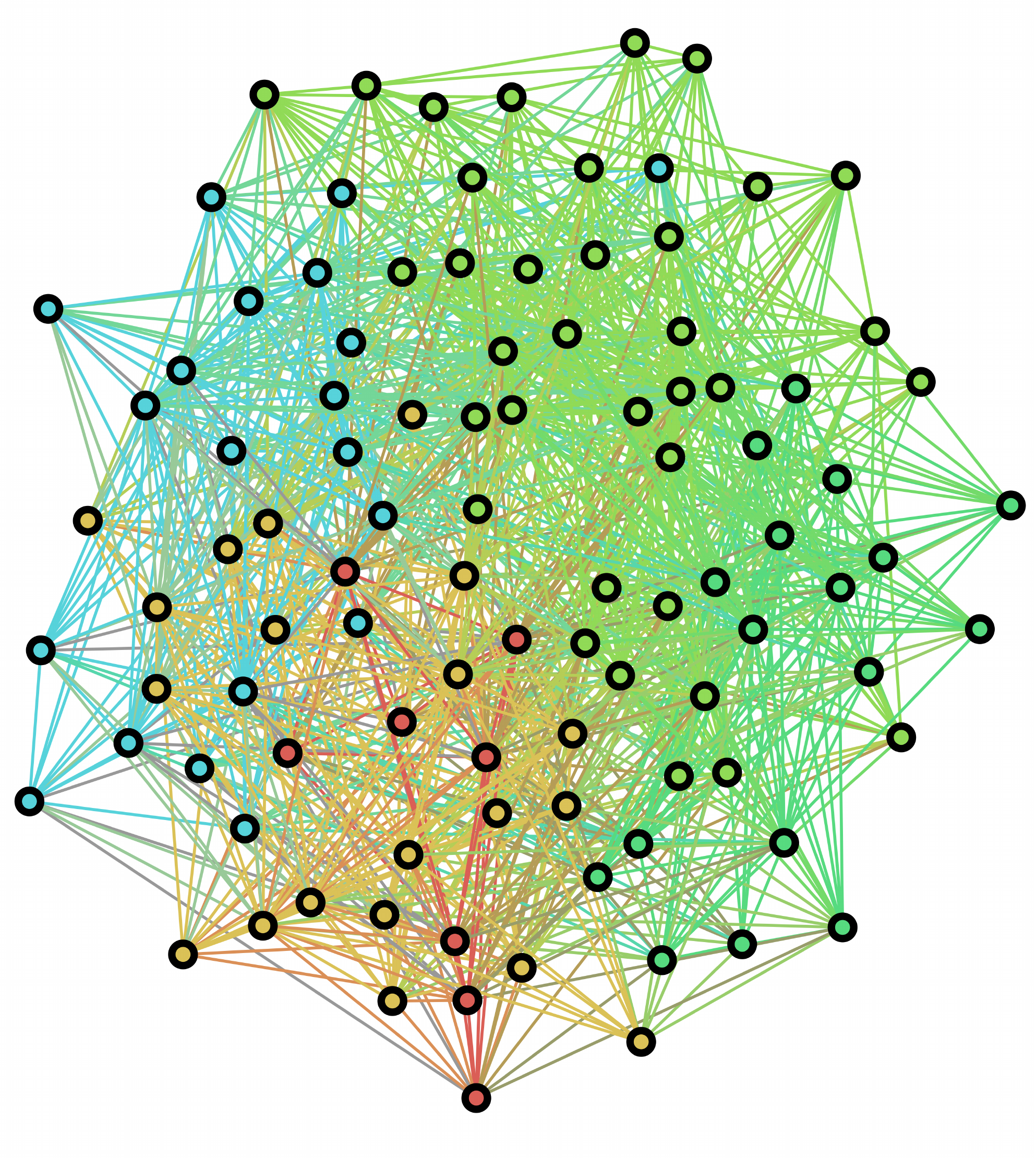}
    \caption{Original Graph}
    \label{fig:original}
  \end{subfigure}
  ~
  \begin{subfigure}[t]{0.4\textwidth}
    \centering
      \includegraphics[width=0.8\textwidth]{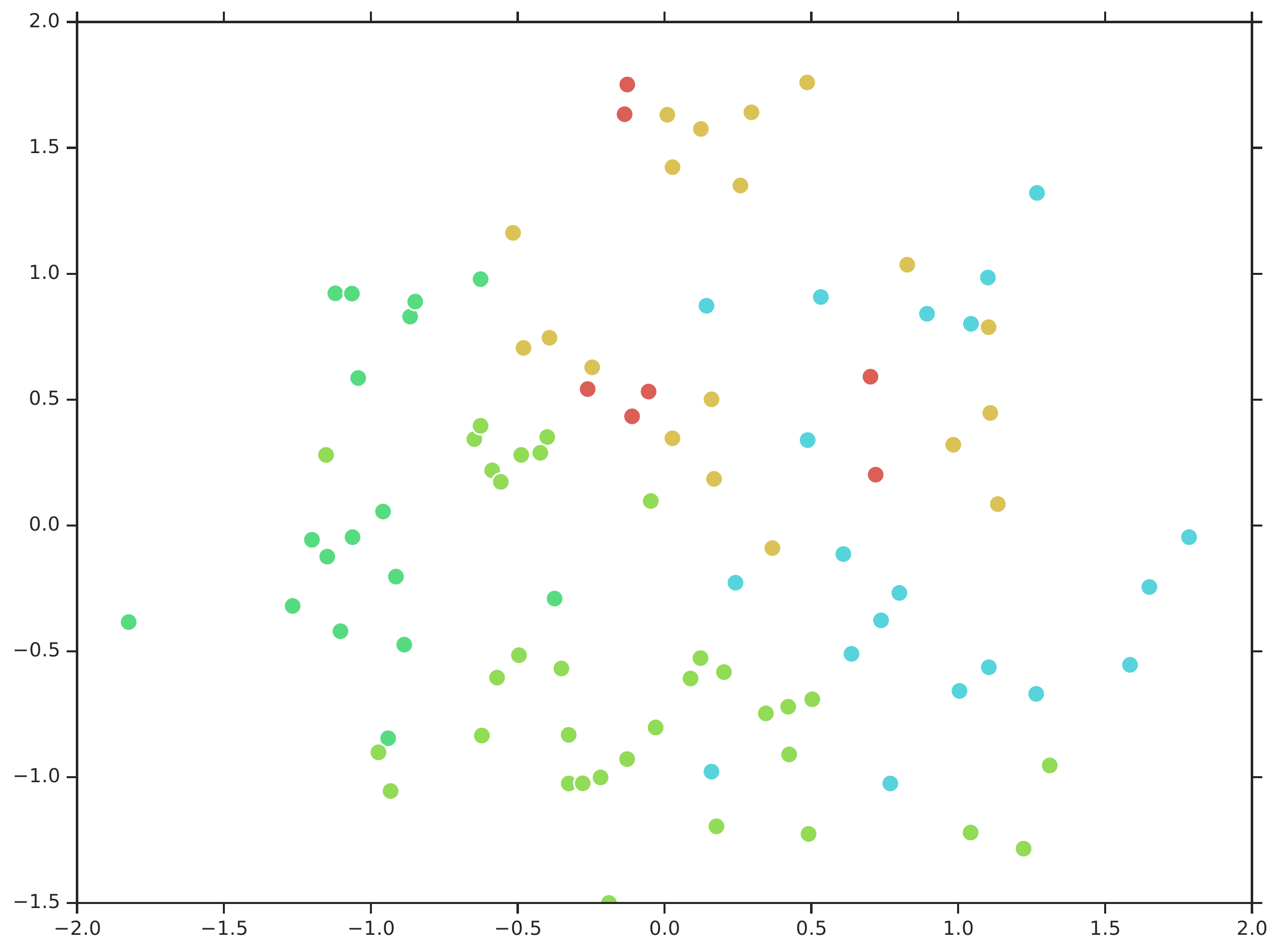}
      \caption{Original Graph Embedding -- using M-NMF}
      \label{fig:original-embeddings}
  \end{subfigure}
\begin{subfigure}[t]{0.4\textwidth}
    \centering
      \includegraphics[width=0.5\textwidth]{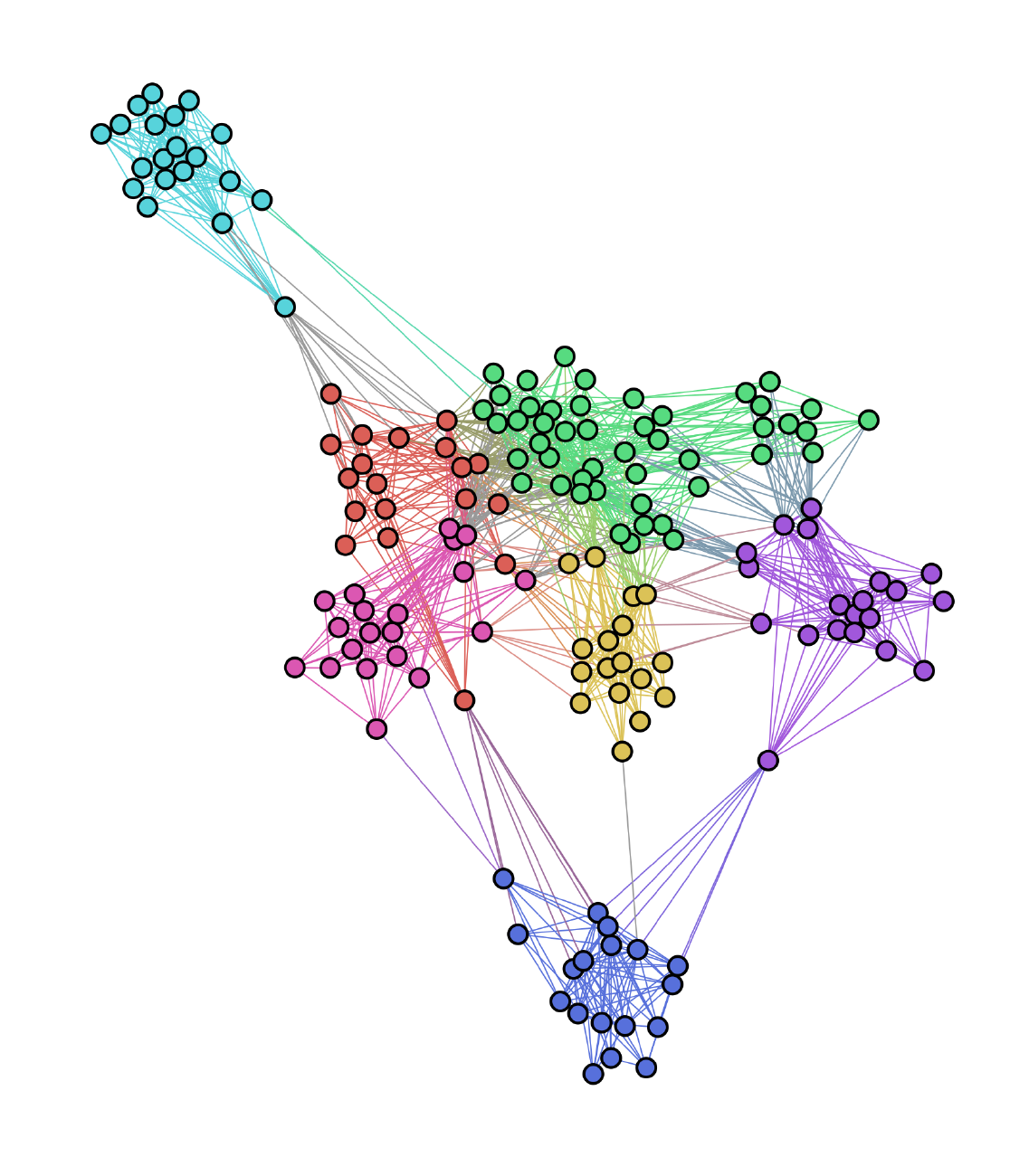}
      \caption{Persona Graph}
      \label{fig:persona}
  \end{subfigure}
  \begin{subfigure}[t]{0.4\textwidth}
    \centering
      \includegraphics[width=0.8\textwidth]{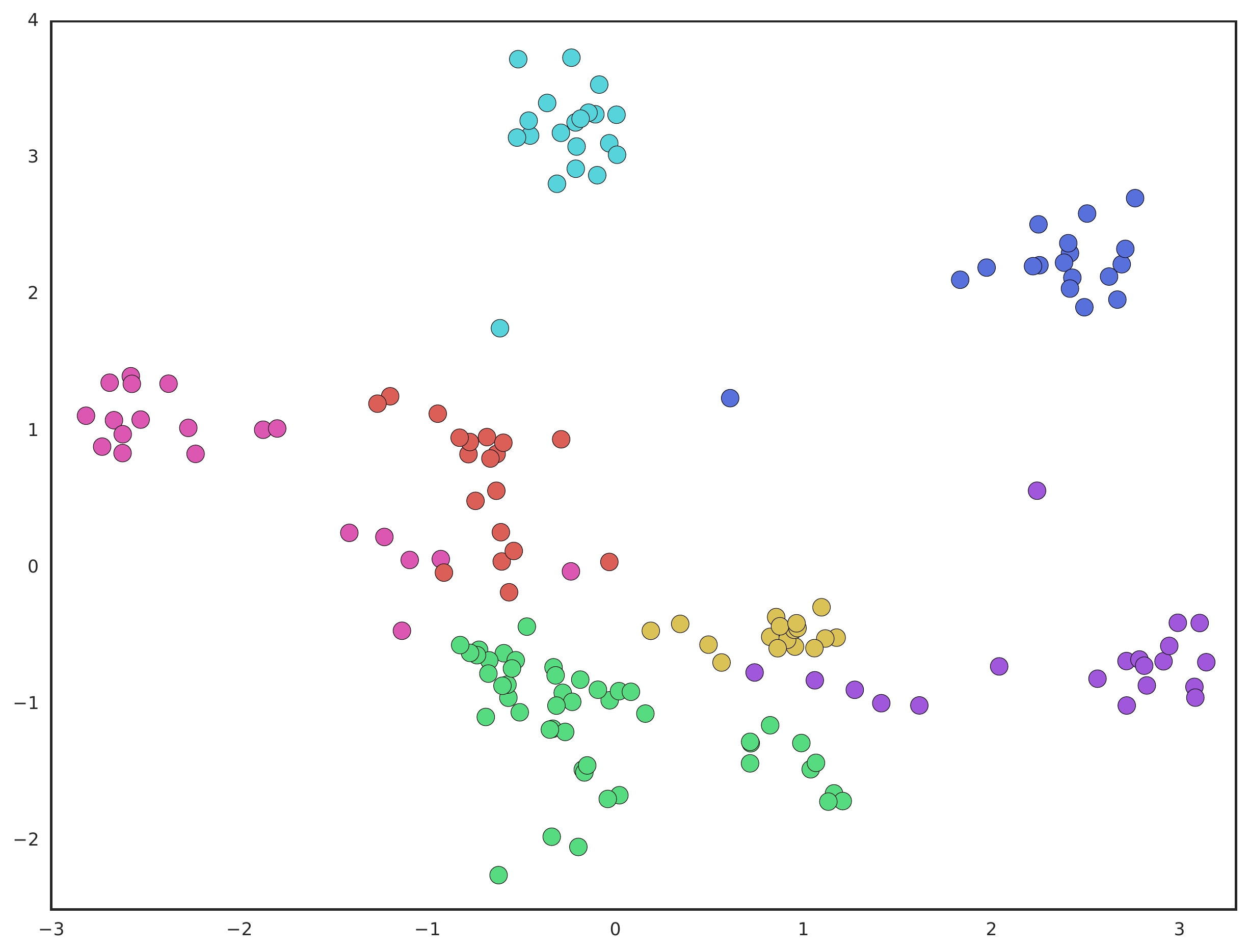}
      \caption{Persona Graph Embedding -- using \splitbrain}
      \label{fig:persona-embeddings}
  \end{subfigure}
  
  \caption{Original graph and Persona graph with corresponding embeddings. Notice how the persona graph community structure is clearer than the one in the original graph and this corresponds to more separated embeddings. The colors in the original and persona graph corresponds to community found by a modularity based algorithm. Left side pictures are used with permission from Epasto et al.~\cite{EpastoKDD2017}. (best viewed in color)}
   \label{fig:persona-vs-original-embeddings} 
\end{figure*}

\subsection{Synthetic graphs}

To gain insight on how our embedding method framework operates, we first provide a visualization of a small synthetic graph. The methodology we use is similar to that of~\cite{EpastoKDD2017}, which we report for completeness. We created a random graph with planted overlapping communities using the Lanchinetti et al~\cite{lancichinetti2009benchmarks} model. We chose this model for consistency with previous ego-net based works~\cite{crgp12,EpastoKDD2017} and because the model replicates several properties of real-world graphs, such as power law distribution of degrees, varying community sizes and the membership of nodes in varying community numbers.
The graph contains $100$ nodes and has $9$ highly overlapping ground-truth communities.

We show the results of the visualization in Figure \ref{fig:persona-vs-original-embeddings}.
First in Figure \ref{fig:original} and Figure~\ref{fig:persona} we show a force-directed layout of the original graph and the corresponding persona graph (using Gephi~\cite{gephi} with the same visualization settings).
The node coloring corresponds to the discovered communities using a non-overlapping community detection that optimizes modularity \cite{blondel2008fast}.
As observed by \citet{EpastoKDD2017} on this dataset, the persona graph has a much clearer community structure, finding 8 of the 9 overlapping communities (only 5 communties around found in the original graph).

We now turn our attention to the embeddings output by our method. In Figure~\ref{fig:original-embeddings}, we show a 2D embedding obtained using M-NMF \cite{wang2017community} with default settings on the original graph, while in Figure~\ref{fig:persona-embeddings}, we show a 2D embedding obtained by our method. As such, it is possible to appreciate how the \splitbrain embeddings more clearly identify the community structure, as the eight communities found are better separated. 
In contrast, the M-NMF embeddings do not show a clear separation for this graph, which has highly overlapping communities.

\subsection{DBLP Co-Authorship Graph}
\label{sec:co-authorship-visualization}
\begin{figure*}[h!]
\centering
  \begin{subfigure}[t]{0.49\textwidth}
    \centering
      \includegraphics[width=\textwidth,trim={2in 2in 2in 2in},clip]{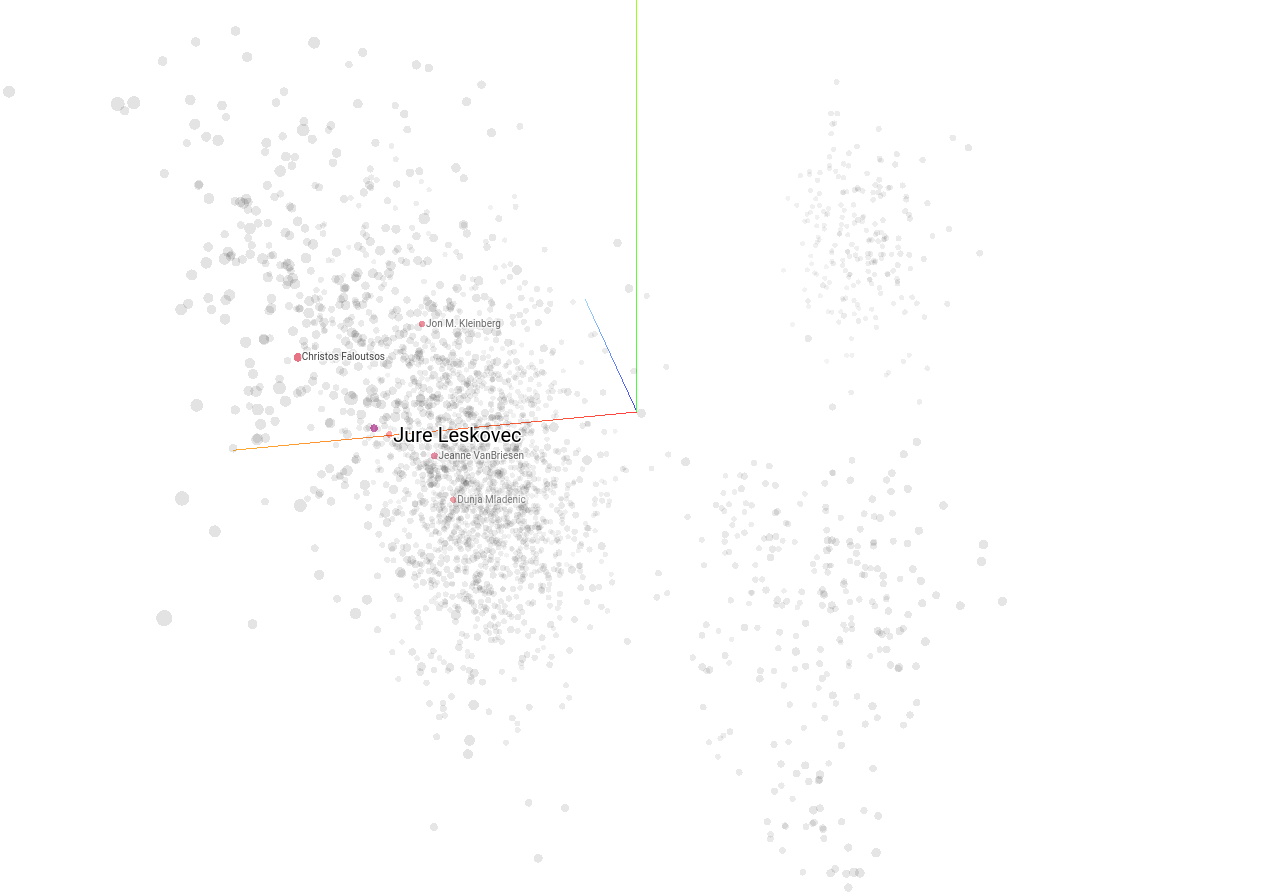}
      \caption{Node2vec -- Highlighted Author}
      \label{fig:original-dblp-leskovec}
  \end{subfigure}
~

  \begin{subfigure}[t]{0.49\textwidth}
    \centering
      \includegraphics[width=\textwidth,trim={2in 2in 2in 2in},clip]{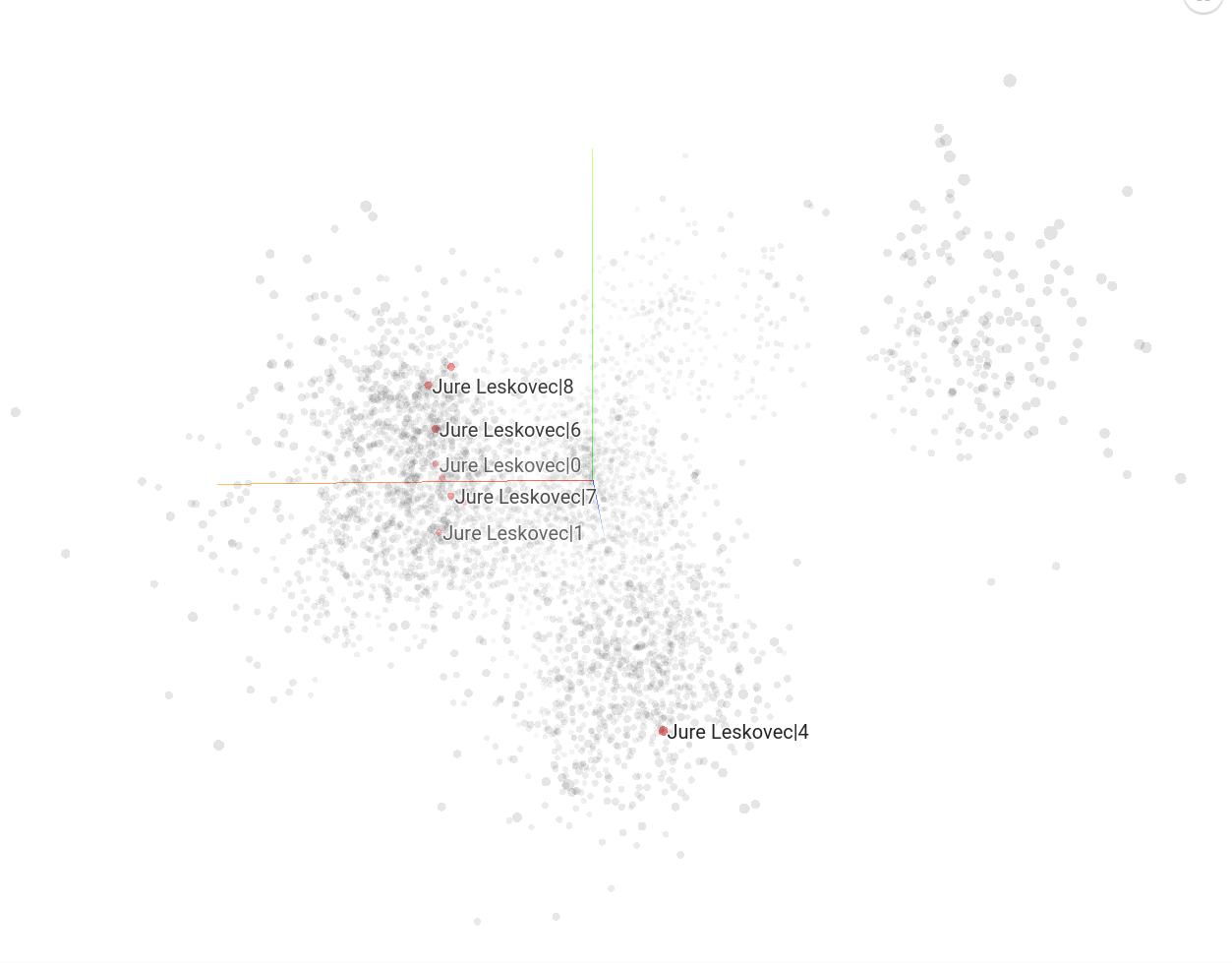}
      \caption{\splitbrain Embeddings -- Highlighted Author}
      \label{fig:persona-dblp-leskovec}
  \end{subfigure}
  
  \caption{Comparison of embedding visualization of a node2vec and \splitbrain in a DBLP co-authorship graph containing authors from $4$ main areas: Data Mining, Machine Learning, Data Base and Information Retrieval. It is possible to observe how the $4$ areas correspond to (more or less) well separated regions in the space in both embedding methods. However, standard embeddings force each author that participates in multiple communities to be represented in a single point, while the persona methods allow author-nodes to be represented by multiple embeddings. We show one such embedding for a prolific author with contributions in both the Data Mining and Machine Learning community. Notice how the node2vec embedding results in the author being embedding in the data mining region only (this is best viewed in color).}\label{fig:persona-vs-original-dblp}
\end{figure*}

We then turn our attention to a real-world co-authorship graph using DBLP data. The \texttt{4area} graph contains co-authorship relationships extracted from papers in $4$ areas of study: Data Mining, Machine Learning, Databases, and Information Retrieval \cite{perozzi2014focused}. 
It is possible to see in Figure~\ref{fig:original-dblp-leskovec} and Figure~\ref{fig:persona-dblp-leskovec}, respectively, a plot of node2vec embeddings and of \splitbrain embeddings. Notice how in Figure~\ref{fig:persona-dblp-leskovec} there are  $4$ areas more or less separated that upon inspection, corresponds to the $4$ different fields. A similar observation is possible for Figure~\ref{fig:original-dblp-leskovec}, representing node2vec embeddings. The key observation in this application scenario is that many authors (in particular the most prolific ones) can contribute to more than one area of study or more than one subarea. However, standard embedding methods force each node to be embedded in only one point in the space, while the \splitbrain method allows to represent a node as a combination of embeddings (in this setting, we obtain $1.58$ personas per node on average). 
In the Figures~\ref{fig:original-dblp-leskovec} and~\ref{fig:persona-dblp-leskovec}, respectively, we highlight the embeddings learned for one such prolific author, \texttt{Jure Leskovec}.  Note how this author has one single embedding obtained by node2vec, and multiple embeddings given in output by \splitbrain.
Upon inspection, we observe that the author is embedded in a data mining region by node2vec, surrounded by other prominent authors in Data Mining, such as \texttt{Christos Faloutsos}.

However, when observing the representations learned through our \splitbrain method, we see that this author has a number of persona representations.
Moreover, many of the personas reflect different sub-groups of coauthors in Data Mining that our node (e.g. one persona corresponds to co-authorship with other students while at CMU).
These personas encode significant portions of the `average' or `global' position, which is captured by node2vec.
Nevertheless, we also see that a persona (\texttt(Jure Leskovec|4)) is now present in the Machine Learning cluster.
This illustrates how the representations from \splitbrain\ allows the node to span both the Data Mining and Machine Learning region of the space, better characterizing the contributions of the node. 
Similar observations hold for other authors.

\section{Related Work}
\label{sect:related-work}
Our work bridges two very active areas of research: ego-net analysis and graph embeddings. As these are vast and fast growing, we will restrict ourselves to
reviewing only the most closely related papers in these two areas.

\subsection{Graph embedding}
These methods learn one embedding per graph node, with an objective that maximizes (minimizes) the product (distance) of node embeddings if they are `close' in the input graph. 
These are most related to our work.
In fact, our work builds on the approach introduced by DeepWalk \cite{deepwalk}, which 
learns node embeddings using simulated random walks.
This idea has been extended to consider node embeddings learned on different variations of random walks \cite{node2vec}, as well as other graph similarity measures \cite{tsitsulin2017verse}, other loss functions \cite{bojchevski2017deep}, or additional information such as edge labels \cite{Chen:2018:ENE:3269206.3269270}.
These node embeddings have been used as features for various tasks 
on networks, such as node classification~\cite{deepwalk}, 
user profiling~\cite{Perozzi:2015:EAP:2740908.2742765}, and link prediction~\cite{node2vec,asymmetric, Zhang:2018:APP:3219819.3219969}.
More recent work in the area has examined preserving the structural roles of nodes in a network \cite{ribeiro2017struc2vec,Tu:2018:DRN:3219819.3220068}, learning embeddings for a graph kernel \cite{rmyeidDDGK}, or proposing attention methods to automatically learn the model's hyperparameters \cite{abu2017watch}.
For more information on node embedding, we direct the reader to a recent survey \cite{chen2018tutorial}.

Moreover, most node embedding methods focus on only learning one representation for each node in the graph.  
Walklets~\cite{walklets} decompose the hierarchy of relationships exposed in a random walk into a family of related embeddings.
However, this is distinctly different from our work, as each node is represented exactly once at each level of the hierarchy.
In addition, the representations are learned independently from each other.
HARP~\cite{chen2017harp} is a meta-approach for finding good initializations for embedding algorithms.
As a by-product, it produces a series of representations that encode a hierarchical clustering.
A number of other works focus on learning community representations, or using communities to inform the node embedding process  \cite{cavallari2017learning,wang2017community,zheng2016node}.
Unlike these works, which focus on aggregating nodes into \emph{less} representations, we focus on \emph{dividing} nodes into more representations.
This allows our approach, \splitbrain, to more easily represent prolific nodes that may have overlapping community membership.

\subsection{Ego-net analysis}

Our work is most closely related to the line of research in social network analysis based on ego-net clustering. 
From their introduction by Freeman~\cite{f82} in 1982, ego-nets or ego-networks are a mainstay of social-network analysis~\cite{b95,dr10,eb05,wf94}. Rees and Gallagher~\cite{DBLP:conf/asunam/ReesG10} jump-started a rich stream of works~\cite{EpastoKDD2017,EpastoVLDB2016,crgp12} that exploit ego-network level clusters to extract important structural information on communities~\cite{fortunato2010community}, to which a node belongs. 
They proposed to partition nodes' ego-net-minus-ego graphs in their connected components to find a global overlapping clustering of the graph.  Coscia et al.~\cite{crgp12} improved over their clustering method by proposing to use a more sophisticated label propagation ego-net partitioning technique. Several authors have since built on such a line of work to improve the scalability and accuracy of ego-net based clustering~\cite{DBLP:conf/icdm/BuzunKAFKTK14,DBLP:conf/bigdataconf/DelisNL16,DBLP:conf/bigdataconf/DelisNL16,EpastoVLDB2016,EpastoKDD2017}, while others have designed ego-net analysis methods that tackle user metadata on top of the ego-net connectivity~\cite{DBLP:conf/nips/McAuleyL12,DBLP:conf/wsdm/YangML14,DBLP:conf/www/LiWC14}.

Mostly related to our work is the recent paper by Epasto el al.~\cite{EpastoKDD2017}, where they introduce the persona graph method for overlapping clustering. They present a scalable overlapping clustering algorithm based on a local ego-net
partition. Their algorithm first creates the ego-nets of all nodes and partition them (in parallel) using any non-overlapping algorithm. These ego-net level partitions are then used to create the persona graph, which is described in more detail in this paper, as this is the basis of our embedding methods. Then, the persona graph is partitioned with another parallel clustering algorithm to obtain overlapping clusters in the original graph.

\section{Conclusions}

We introduced \splitbrain, a novel graph embedding method that builds on recent advances in ego-network analysis and overlapping clustering. In particular, we exploited the recently introduced persona graph decomposition to develop an embedding algorithm that represents nodes in the graph with multiple vectors in a principled way. 
Our experimental analysis shows strong improvements for the tasks of link prediction and visual discovery and exploration of the community membership of nodes. 

Our method draws a connection between the rich and well-studied field of overlapping community detection and the more recent one of graph embedding which we believe may result in further research results.
As future work we want to explore more in this direction, focusing on the following challenges: (1) exploiting embeddings for overlapping clustering; (2) studying the effect of this method on web-scale datasets; (3) developing theoretical results on this method; (4) applying our embeddings for classification and semi-supervised learning tasks; and (5) developi

\balance

\bibliographystyle{ACM-Reference-Format}
\bibliography{references}

\end{document}